\documentclass{nature}

\newcommand{\newsrc}{GCRT~J1745$-$3009}

\newcommand{\mjybm}{\mbox{mJy~beam${}^{-1}$}}
\newcommand\arcdeg{\mbox{$^\circ$}}%
\newcommand\arcmin{\mbox{$^\prime$}}%
\newcommand\arcsec{\mbox{$^{\prime\prime}$}}%
\newcommand{\apj}{Astrophys.\ J.}
\newcommand{\aj}{Astron.\ J.}
\newcommand{\nat}{Nature}

\title{A powerful bursting radio source towards the Galactic Center}

\author{Scott D.~Hyman$^{1}$, T.~Joseph W.~Lazio$^2$, Namir E.~Kassim$^2$,
  Paul S.~Ray$^3$ \& Craig B.~Markwardt$^4$
 \& Farhad Yusef-Zadeh$^5$}

\begin{document}

\maketitle

\begin{affiliations}
 \item Department of Physics and Engineering, Sweet Briar College, Sweet Briar, VA 24595
\item Naval Research Laboratory, Code 7213,
	Washington, DC 20375-5320
\item E. O. Hulburt Center for Space Research,
 Naval Research Laboratory, Washington, DC 20375
 \item Laboratory for High Energy Astrophysics, NASA Goddard Space Flight
Center, Greenbelt, Maryland 20771
\item Northwestern University, Department of Physics and Astronomy,
  Evanston, Illinois 60208
\end{affiliations}

\begin{abstract}
Transient astronomical sources are typically powered by compact objects and usually signify
highly explosive or dynamic events\cite{cordes04}. While radio astronomy 
has an impressive record of obtaining high time 
resolution observations\cite{hkwe03}, usually it is achieved in quite narrow
fields-of-view.  Consequently, the dynamic radio sky is poorly sampled, in 
contrast to the situation in the X- and $\gamma$-ray bands in which 
wide-field instruments routinely detect transient sources\cite{blrs01}.  Here we 
report a new transient source, \newsrc, detected in 2002 during a moderately 
wide-field radio transient monitoring program of the Galactic center 
(\hbox{GC}) region\cite{hyman02,hyman03} at 0.33 GHz. The characteristics of its bursts are 
unlike those known for any other class of radio transient. If 
located in or near the \hbox{GC}, its brightness temperature
($\sim10^{16}\,\mathrm{K}$)
and the implied energy density within \newsrc\
vastly exceeds that observed in most other classes of radio 
astronomical sources\cite{readhead94}, and is consistent with
coherent emission processes\cite{melrose02} rarely observed. We conclude that \newsrc\ is the first 
member of a new class of radio transient sources, the first of 
possibly many new classes to be identified through current and 
upcoming radio surveys\cite{kassim04}. 
\end{abstract}

\newsrc\ is located approximately 1.25\arcdeg\ south of
the Galactic center (\hbox{GC}) and is
notable for a series of $\sim1$~Jy bursts, each with a duration of $\sim10$~min.,
and occurring at apparently regular intervals of~1.27~hr. The variability of \newsrc\ is shown in the light
curve of Figure~1, and the average burst light curve is shown in Figure~2.
The light curves appear to be
similar in shape, although the missing data during the first, second,
and third bursts hinders a comprehensive comparison. 
 \newsrc\ is located near the supernova remnant,
SNR~359.1$-$00.5$^{[}$\cite{reich84}$^{]}$, and other prominent sources\cite{yusefzadeh87,gray91}, as
shown in Figure~3.
\newsrc\ was detected using the Very Large Array radio telescope operating at a
frequency of 0.33 GHz.

\newsrc\ is not detected between bursts with a $5\sigma$ upper limit
of 75 mJy, determined by imaging the entire observation with the
bursts removed. We also do not detect the source in 0.33~GHz, $\sim
1$-hr GC monitoring observations made earlier in 2002 and afterwards
in~2003; the $5\sigma$ upper limit for detection in a bursting state
is $\sim 250$~mJy with 5-min integrations, and in a quiescent state is
$\sim 50$~mJy. Images made from three 6-hr observations in~1996 and
1998 have similar upper limits, and the combination of these images\cite{nlkhlbd04}
has a 15~mJy upper limit for quiescent emission.

The magnitude of errors in radio astronomical images typically
increases with distance from the center of the image. \newsrc\ is
located only 14\arcmin\ from the image center compared to the
$\sim$3\arcdeg\ field-of-view, and therefore, together with its
detection at multiple frequencies around 0.33~GHz and in both circular polarizations, we consider
the evidence to be very
strong that the source is real.
The bursts show no significant frequency dependence and no molecular-line
masers are known to emit near 0.33~GHz; therefore, the lack of a frequency dependence suggests
that \newsrc\ is not a maser.

\newsrc\ is unresolved in our observation. If we constrain its size to be less than 
$c\tau$, with $\tau \sim 2$~min.\ taken to be the decay time of the 
$\sim1$~Jy bursts, the energy density within the source as measured by the 
brightness temperature is $\sim10^{12}\,\mathrm{K}\,(D/70\,\mathrm{pc})^{2}$,
where $D$ is the distance to the source.  If the transient source is at the 
\hbox{GC}, $\sim$8.5 kpc distant, its brightness temperature far 
exceeds $10^{12}$~K, the upper limit for incoherent synchrotron 
radiation\cite{readhead94} produced by relativistic electrons gyrating in a magnetic 
field, and therefore its emission is likely coherent. 

In principle, \newsrc\ could be located $< 70$~pc from us, in which 
case it could be either a coherent or an incoherent emitter. 
Known and hypothesized classes of ``local'' ($D < 70$~pc)
sources that show flare activities include dwarf M-type (dMe) stars,
brown dwarfs, and extrasolar planets. 
dMe flare stars emit coherent bursts produced through electron cyclotron 
maser emission. 
The bursts from flare stars 
show some similarities to the light curve of \newsrc, but in contrast, they
are detected in only one circular polarization
at low frequencies (e.g., at 0.43~GHz for AD Leo and YZ CMi)\cite{bbdd90}. Bursts of such highly circularly polarized
radio emission are also predicted, 
by analogy to the giant planets in the solar system, from extrasolar
giant planets
\cite{fdz99,lazio04}. However, no detections have been made
in searches for such emission from known extrasolar planets at 0.33~GHz and 1.5~GHz, at comparable
or better sensitivity limits to what we report here\cite{bastian00}.  We conclude
that \newsrc\ is not likely to be a dMe flare star or an extrasolar planet.

Brown dwarfs also emit flares, apparently as a result of processes 
involving high magnetic fields\cite{guedel02}.
Four infrared sources 
detected in the {\it{Two Micron All Sky Survey}}\cite{cutri03} that lie within
the $\sim10\arcsec$ uncertainty of the transient's radio
position could possibly be brown dwarfs. However, their distances are unknown and their spectral
colors, to the extent measured, are inconsistent with those of brown
dwarfs\cite{bkb+02}.

Few radio 
flares from brown dwarfs have been detected\cite{berger01,berger02,putman03},
and none at low frequencies\cite{bastian00}.
Like the bursts from \newsrc, the observed flares have time scales 
of minutes, and, for two of the brown dwarfs, LP944-20 and 
2MASS~0036$+$18, they also have faster decay than rise times.  In 
contrast to \newsrc, however, no regular flaring pattern is evident for
any of these sources, and
they were observed only at high frequencies 
since emission is predicted to be self-absorbed at lower 
frequencies\cite{dulk82}. The extent to which flares from brown 
dwarfs are self-absorbed at low frequencies is somewhat uncertain, 
though, since flares from only two objects
(LP944-20 and DENIS~1048-3956) have been observed
at multiple frequencies simultaneously, and 
these frequencies were much higher (4.8 and 8.5 GHz) than 0.33 GHz
at which we detect \newsrc.
Furthermore, while, in contrast to the low-frequency bursts
from \newsrc, the observed flares from brown dwarfs
at high frequencies are significantly polarized 
(30--70$\%$), the
degree of polarization may be significantly reduced at 
low frequencies where self-absorption can become 
large\cite{bruggmann90}. 

We conclude, therefore, that we cannot rule out that \newsrc\ is a flaring brown dwarf. However, its detection at a low frequency, its
regular flaring pattern, and significant detection in both
circular polarizations, are all novel features in comparison to known
characteristics of brown dwarfs. Hence, there is no compelling evidence
for an identifcation as such.

While \newsrc\ is conceivably a ``local'' radio source belonging to one of the classes of
objects considered above,
it is much more likely that it is located significantly
further from us.  Assuming even a uniform distribution of transients,
and not the vastly increasing spatial density of all astronomical
objects toward the \hbox{GC}, the relative spatial volume covered by
our wide-field observations results in an extremely small probability
($6 \times 10^{-7}$) that \newsrc\ is located within 70~pc.

Many transient radio sources are also detectable at X-ray and $\gamma$-ray
wavelengths. We have analyzed a serendipitous pointed observation about
32\arcmin\ from \newsrc\ in the {\it{Rossi X-Ray Timing Explorer}} (RXTE) archive 
that runs between the third and fourth burst, though 
overlapping neither. No variable X-ray (2--10 keV) emission was seen and we
place a conservative 25 mcrab  upper limit on the flux of any
interburst X-ray counterpart emission (1 mcrab $\approx$ $2\times10^{-11}$ erg cm$^{-2}$
s$^{-1}$).
One of us (C.B.M.)
performs regular scanning observations of the Galactic Bulge with
RXTE. The closest of these were on 2002 September 25 and October 2.
In both of these observations, 3$\sigma$ upper limits on the X-ray
flux from the source are 6 mcrab.  A search of all of the Bulge scans
starting 1999 February finds one scan, on~2003 July~3, in which a 15
mcrab (6.5$\sigma$) outburst was detected. However,  
possible confusion with other X-ray sources in the field-of-view ($\sim 30\arcmin$)
prevents a conclusive identification.
Similarly,
the $\gamma$-ray source 3EG~J1744$-$301 was detected
near the position of \newsrc\ during the 1990's by the {\it{Energetic Gamma
Ray Experiment Telescope}}, but the positional error on that source is also large
($\sim 20\arcmin$) and the source is in a highly confused region with numerous diffuse and discrete sources of $\gamma$-ray emission\cite{mh98}.

We next consider the possibility that the bursts from \newsrc\ could be
relativistically beamed toward us as is the case for microquasars, which are
accreting black holes
in binary systems that occasionally power radio-bright relativistic jets
(e.g., GRS~1915+105, whose
apparent superluminal motion has a Lorentz factor of
$\gamma\sim5$)\cite{marscher02,fender99}. While it is conceivable that relativistic
beaming is responsible for the high calculated brightness temperature, the light curve
of \newsrc\ does not resemble that for known microquasars or other sources of jet
emission, most of which exhibit a fast rise and a slower decay and much longer time scales\cite{blrs01}. In addition, the apparent lack of a bright X-ray counterpart to the bursts argues strongly against accretion as the power source for the bursts.

Next, we consider radio pulsar origins for the source.  A 77-min rotation period radio pulsar is excluded because the rotational energy loss rate from such a pulsar is insufficient to power the radio emission unless the magnetic field is extreme ($> 10^{18}$ G) or the distance is unreasonably small ($< 0.5$ pc).  Another conceivable option is that the 77 minutes is an orbital period and the outbursts are flux variations as a function of orbital phase similar to the pulsar PSR J0737$-$3039B$^{[}$\cite{lbk+04}$^{]}$.  This scenario does not explain the transient behavior from the source or the lack of interburst emission, and favors a distance of order 1 kpc or less.  

One other class of sources to consider are magnetars, neutron stars with immense ($10^{14} - 10^{15}$ G) magnetic fields whose radiation is powered by field decay\cite{woods05}.  Coherent emission from the magnetosphere of a magnetar addresses the energy budget difficulties seen in the pulsar models.  An investigation of whether magnetars could produce the observed emission timescales and transient behavior is underway (K.S.~Wood, P.S.R., S.D.H., T.J.W.L., \& N.E.K. in preparation). 

\begin{figure}
\centerline{\includegraphics{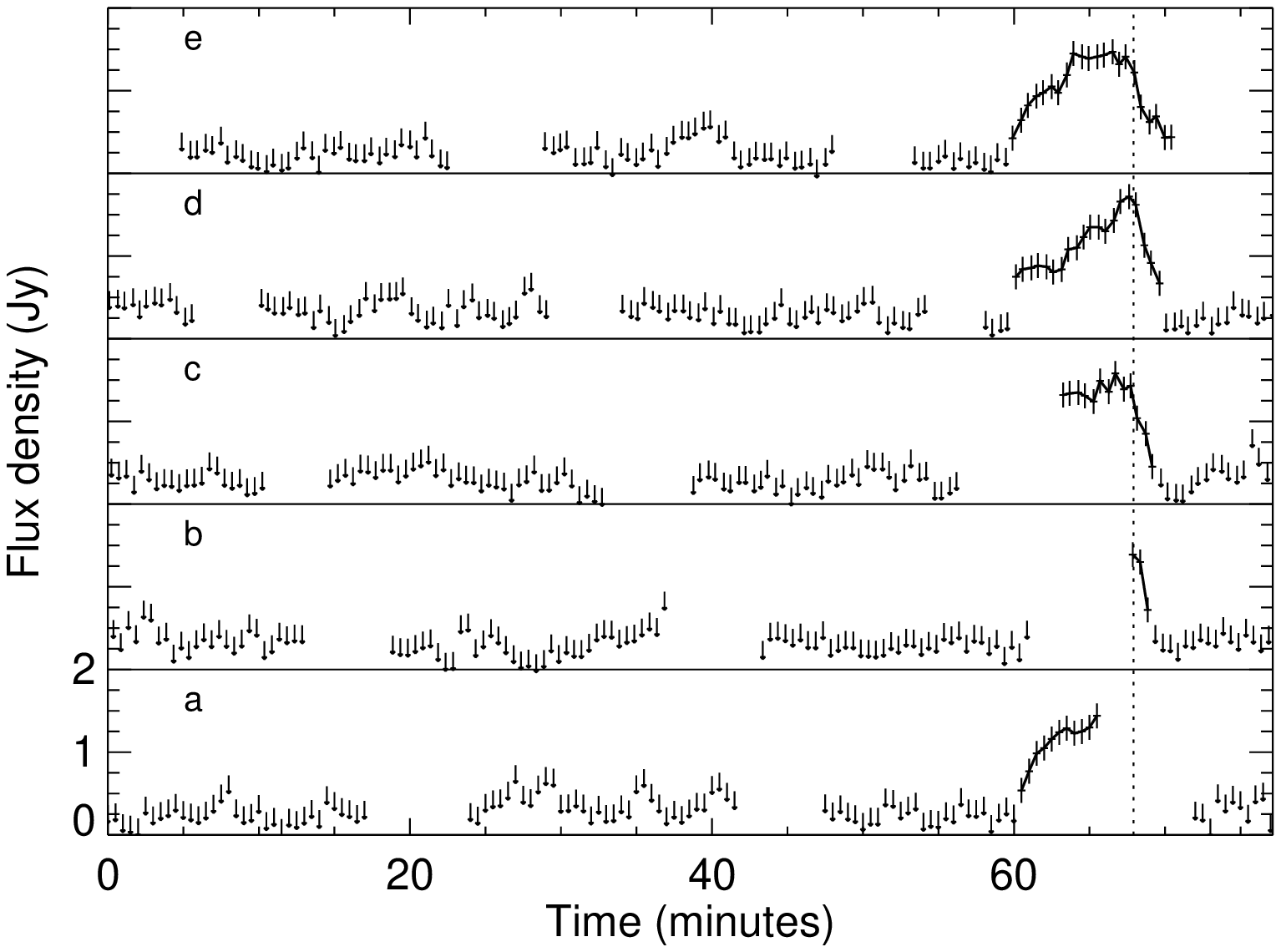}}
\caption[]{The five detected bursts from the radio transient source,
\newsrc.  The observation is continuous, with the time axis folded at multiples of 77.13 min.
{\bf{a}}, The first interval of the observation beginning at 20:50:00 on
30 September 2002 (TAI).
{\bf{b}}, The second interval. {\bf{c}}, The third interval. {\bf{d}}, The fourth interval.
{\bf{e}}, The fifth interval.
The points connected by the heavy line are the detections in 30-s samples with typical error bars of 0.15 Jy shown. The arrows are
3$\sigma$ upper limits for nondetections between bursts; no evidence of
quiescent emission is found. Fluctuations of nearby sources are consistent
with the noise level. The dashed vertical line is positioned at the fitted peak
(see Figure~2) of the fourth burst as a reference.
Note that several gaps in the data, including during the first three bursts,
are due to radio frequency interference or when the phase calibrator was
imaged. No anomalous behavior is seen for the calibrator.
The 0.33~GHz, 7-hr observation was obtained
with the CnB-configuration of the Very Large Array.
The bandpass consists of
thirty-one, 97~kHz wide channels for each of
two intermediate frequencies (IF, 321.56 and~327.50~MHz). Both circular
polarizations were imaged, but linear
polarization measurements are not available for this observation.
No circular polarization was detected (15$\%$, $5\sigma$ upper limit)}.
\label{fig:lightcurve}
\end{figure}

\begin{figure}
\centerline{\includegraphics{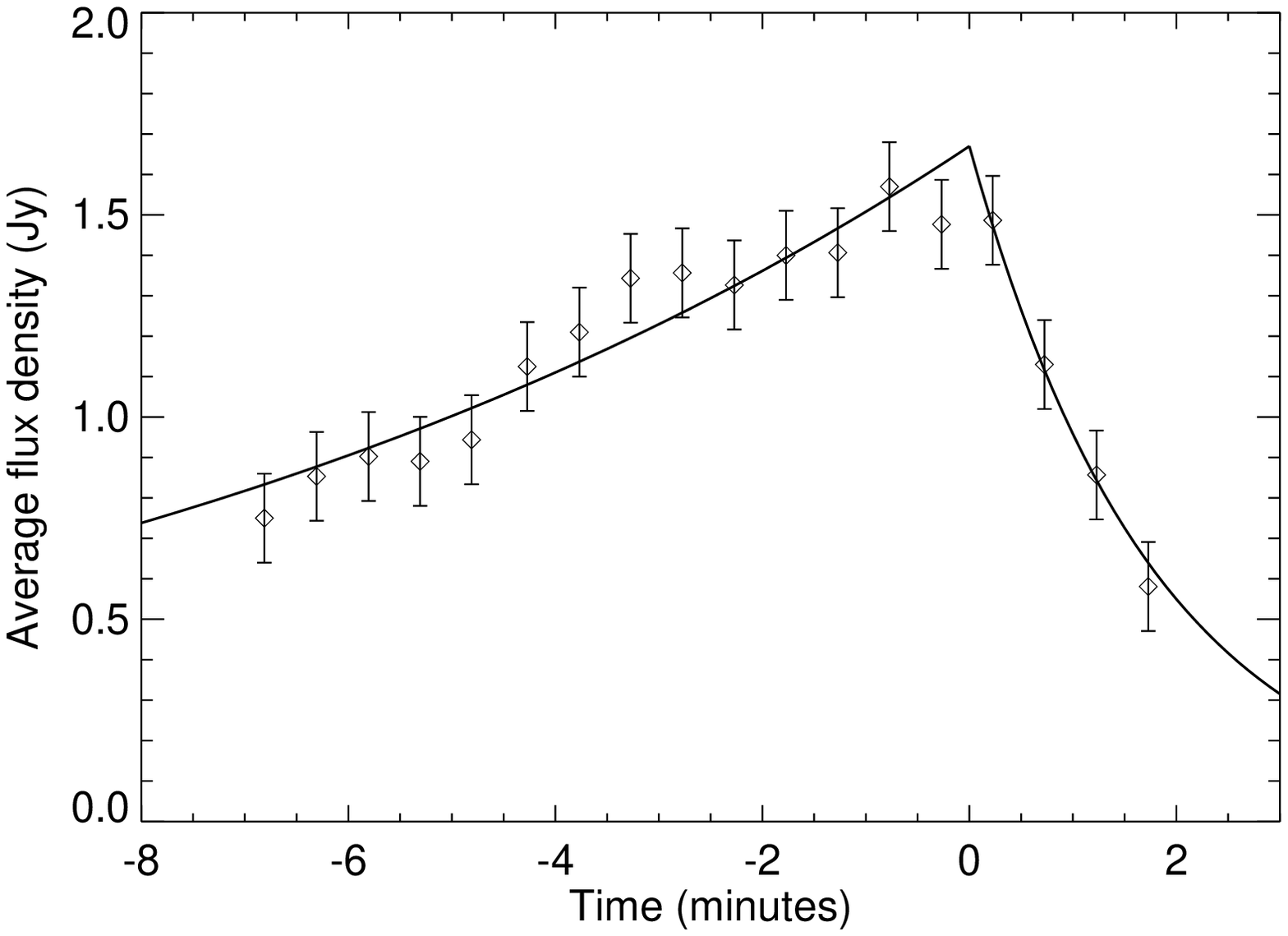}}
\caption[]{Average light curve of \newsrc\ derived from the third, fourth, and fifth bursts.  The peak times of the bursts, $t_{o}$, were determined for
all but the second, most incompletely
sampled burst,
from exponential fits
 to their rise ($Ae^{(t-t_{o})/\tau_{1}}$) and decay
  ($Ae^{-(t-t_{o})/\tau_{2}}$) and have been subtracted from the light curve of each
in order to construct the average. (While the first burst was included in the fit, it 
also is largely undersampled and therefore not included in the average light curve.)
The peak of the average light curve is arbitrarily placed at 0~min. Typical error bars
($\pm0.11$~Jy) are shown.
The amplitude, $A$, and rise and decay
  time constants, $\tau_{1}$ and $\tau_{2}$, were constrained to be identical in the
fits, yielding $A = 1.67 \pm 0.05$~Jy, $\tau_{1} = 9.9 \pm 0.7$~min, and
   $\tau_{2} = 1.9 \pm 0.5$~min. 
The difference between the peak times derived for the first and third bursts is
$2\times77.3 \pm 0.2$ min., the third and fourth $77.6 \pm 0.3$
min., and the fourth and fifth $76.5 \pm 0.3$ min. A separate fit (solid curve) was
made to the average light curve and yields parameters consistent with those listed above.}
\label{fig:ltcurveoverlap}
\end{figure}

\begin{figure}
\centerline{\includegraphics{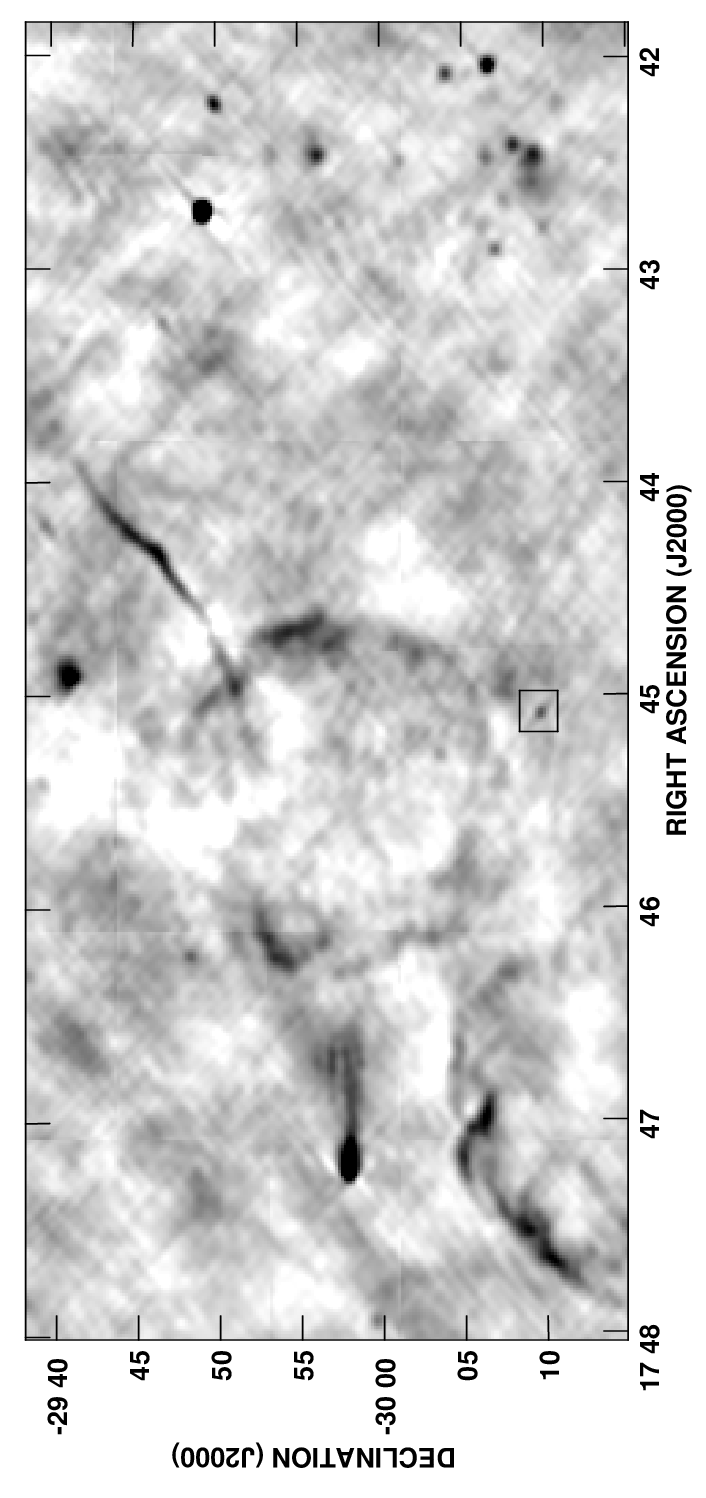}}
\caption[]{0.33 GHz radio image of the transient source, \newsrc, and the surrounding region
$\sim1\arcdeg$ south of the Galactic center.  
\newsrc\ is located at (J2000) R.A. $17^{h} 45^{m} 05 \pm 0.8^{s}$,
Dec. $-30\arcdeg 09\arcmin 52 \pm 10\arcsec$, indicated by the small box below the
$\sim20\arcmin$~diameter shell of the supernova remnant, SNR~359.1$-$00.5. 
Other sources in the image include 
the sources to the west which are part of {\it{Sgr~\hbox{E}}}, the
linear feature, {\it{The Snake}}, to the north, and {\it{The Mouse}} to the northeast of \newsrc.
The sensitivity and
resolution of the image are 15~\mjybm and $48\arcsec \times 39\arcsec$,
respectively. We searched
the entire $3\arcdeg$ field-of-view for other intra-observation transients and
variable sources and found none. Note that \newsrc\ appears as only a 100 mJy
source here since it is averaged over the five short $\sim$1~Jy
bursts and a total of $\sim$6-hr of nondetections between bursts.
Nondetections of quiescent and bursting emission at 0.33 GHz for other epochs are described in the text. We also
do not detect quiescent emission in a 1.4~GHz observation
from 2003 January, with a detection threshold of~35~mJy.  However, if
\newsrc\  has a steep spectrum, as found for other radio
transients (e.g., $\alpha = -1.2$, $S_{\nu} \propto \nu^\alpha$ for the
{\it{Galactic Center Transient}}\cite{zhaoetal92}) and coherent emitters such as radio pulsars ($\alpha \sim -1.7$), it likely could
have decayed significantly below the 2003 January detection threshold.
Also, while no bursts were detected within the 1.4~GHz observations,
these data were comprised of seven~3-min.\ ``snapshots'' taken every
hour; given its 1.25~hr recurrence interval, \newsrc\ could have
been missed easily in a bursting state. Similarly, most of the
0.33~GHz nondetection observations from 2002 and 2003 were only about an hour or
less in duration, and therefore it is possible that \newsrc\ was and
still is detectable in a bursting state.}
\label{fig:SNRimage}
\end{figure}


\begin{addendum}

 \item 
The Very Large Array (VLA) is operated by the National Radio Astronomy
   Observatory (NRAO), which is a facility of the
   National Science Foundation operated under cooperative agreement by
   Associated Universities, Inc. Basic research in radio astronomy is supported at Sweet Briar College by funding from the
   Jeffress Memorial Trust and Research Corporation. Basic research in
   radio and X-ray astronomy at the NRL is supported by the Office of Naval
   Research. S.D.H. thanks J.~Neureuther and M.~Lazarova for their assistance
   in the Galactic center transient monitoring program. The authors thank
   Drs. Deepto Chakrabarty, Charles Dermer, William Erickson, 
   Ronald Remillard, Kurt Weiler, and Kent Wood for helpful discussions.

\end{addendum}



\begin{thebibliography}{1}
\bibitem{cordes04} Cordes, J.~M., Lazio, T.~J.~W. \& McLaughlin, M.~A. The dynamic
	radio sky. {\it{astro-ph}}, http://arxiv.org/abs/astro-ph/0410045 (2004).
\bibitem{hkwe03} Hankins, T.~H., Kern, J.~S., Weatherall, J.~C. \& Eilek, J.~A.
	Nanosecond radio bursts from strong plasma turbulence in the Crab pulsar. 
        {\it{\nat}} {\bf{422}}, 141-143 (2003). 
\bibitem{blrs01} Bradt, H., Levine, A., Remillard, R. \& Smith, D.~A. X-ray transients
	monitored by the All-Sky Monitor on RXTE: A tabulation. in
	{\it{AIP Conf.~Proc.: X-ray Astronomy: Stellar Endpoints, AGN, and the
	Diffuse X-ray Background}} (eds. White, N.~E., Malaguti, G. \&
	Palumbo, G.~G.~C.) {\bf{599}}, 35-52 (2001).
\bibitem{hyman02} Hyman, S.~D., Lazio, T.~J.~W., Kassim, N.~E.
        \& Bartleson, A.~L. Low-frequency radio transients in the galactic
         center. {\it{\aj}} {\bf{123}}, 1497-1501 (2002).
\bibitem{hyman03} Hyman, S.~D., Lazio, T.~J.~W., Kassim, N.~E., Nord, M.~E.
        \& Neureuther, J.~L. A search for radio transients at 0.33 GHz in the
        GC. {\it{Astron. Nachr.}} {\bf{324(S1)}}, 79-83 (2003).
\bibitem{readhead94} Readhead, A.~C.~S. Equipartition brightness temperature and the
	inverse Compton catastrophe. {\it{\apj}} {\bf{426}}, 51-59 (1994).
\bibitem{melrose02} Melrose, D.~B. Coherent emission in AGN: a critique.
	{\it{Publ.~Astron.~Soc.~Aust.}} {\bf{19}}, 34-38 (2002).
\bibitem{kassim04} Kassim, N.~E. The long wavelength array. {\it{http://lwa.nrl.navy.mil}} (2004).
\bibitem{reich84} Reich, W. \& F\"urst, E. G357.7 + 0.3 and G359.1 - 0.5 - two
	shell-type supernova remnants in the galactic centre region.
	{\it{A\&AS}} {\bf{57}}, 165-167 (1984). 
\bibitem{yusefzadeh87} Yusef-Zadeh, F. \& Bally, J. A non-thermal axially symmetric radio wake
	towards the galactic centre. {\it{\nat}} {\bf{330}}, 455-458 (1987).
\bibitem{gray91} Gray, A.~D., Cram, L.~E., Ekers, R.~D. \& Goss, W.~M. A filamentary radio source
	near the galactic centre. {\it{\nat}} {\bf{353}}, 237-239 (1991).
\bibitem{nlkhlbd04} Nord, M.~E. {\it{et al.}} High-resolution, wide-field
        imaging of the galactic center region at 330 MHz. {\it{\aj}}
         {\bf{128}}, 1646-1670 (2004).
\bibitem{bbdd90} Bastian, T.~S., Bookbinder, J., Dulk, G.~A. \&
	Davis, M. Dynamic spectra of radio bursts from flare stars.
	{\it{\apj}} {\bf{353}}, 265-273 (1990).
\bibitem{fdz99} Farrell, W.~M., Desch, M.~D. \&  Zarka, P. On the possibility of
	coherent cyclotron emission from extrasolar planets. {\it{J.~Geoph.~Res.}} 
	{\bf{104}}, 14025-14032 (1999). 
\bibitem{lazio04} Lazio, T.~J.~W. {\it{et al.}} The radiometric Bode's Law and
	extrasolar planets. {\it{\apj}} {\bf{612}}, 511-518 (2004).
\bibitem{bastian00} Bastian, T.~S., Dulk, G.~A. \& Leblanc, Y. A search for radio emission
	from extrasolar planets. {\it{\apj}} {\bf{545}}, 1058-1063 (2000).
\bibitem{guedel02} Guedel, M. Stellar radio astronomy: probing stellar atmospheres from
	protostars to giants. {\it{Annu. Rev. Astron. Astrophys.}} {\bf{40}}, 217-261 (2002).
\bibitem{cutri03} Cutri, R.~M. {\it{et al.}} The 2MASS all-sky catalog of point sources.
	University of Massachusetts and Infrared Processing and Analysis Center,
	IPAC/California Institute of Technology, (2003).	
\bibitem{bkb+02} Burgasser, A.~J. {\it{et al.}} The spectra of T dwarfs. I. Near-infrared
 	data and spectral classification. {\it{\apj}} {\bf{564}}, 421-451 (2002).
\bibitem{berger01} Berger, E. {\it{et al.}} Discovery of radio emission 
	from the brown dwarf LP944-20. {\it{\nat}} {\bf{410}}, 338-340 (2001).
\bibitem{berger02} Berger, E. Flaring up all over-radio activity in rapidly rotating
	late M and L dwarfs. {\it{\apj}} {\bf{572}}, 503-513 (2002).
\bibitem{putman03} Putman, M.~E. \& Burgasser, A.~J. Radio emission from late-type dwarfs:
	quiescent emission and a spectacular radio flare from the M9 DENIS 1048-3956.
	{\it{Bull. Am. Astron. Soc.}} {\bf{35}},
	http://www.aas.org/publications/baas/v35n5/aas203/1159.htm (2003).
\bibitem{dulk82} Dulk, G.~A. \& Marsh, K.~A. Simplified expressions for the gyrosynchrotron
	radiation from mildly relativistic, nonthermal and thermal electrons. {\it{\apj}}
	{\bf{259}}, 350-358 (1982).
\bibitem{bruggmann90} Bruggmann, G. \& Magun, A. Temporal and spectral characteristics of
	the circular polarization of solar microwave bursts. {\it{A\&A}} {\bf{239}}, 347-355 (1990).
\bibitem{mh98} Mayer-Hasselwander, H.~A. {\it{et al}}. High-energy gamma-ray emission
	from the Galactic Center. {\it{A\&A}} {\bf{335}}, 161-172 (1998).
\bibitem{marscher02} Marscher, A.~P. {\it{et al.}} Observational evidence for the
	accretion-disk origin for a radio jet in an active galaxy. {\it{\nat}}
	{\bf{417}}, 625-627 (2002).
\bibitem{fender99} Fender, R.~P. {\it{et al.}} Merlin observations of relativistic
	ejections from GRS~1915+105.
	{\it{Mon.~Not.~R.~Astron.~Soc.}} {\bf{304}}, 365-876 (1999).
\bibitem{lbk+04} Lyne, A.~G. {\it{et al.}} A double-pulsar system: a rare laboratory
	for relativistic gravity and plasma physics. {\it{Science}} {\bf{303}},
	1153-1157 (2004).
\bibitem{woods05} Woods, P.~M. \& Thompson, C. Soft gamma repeaters and anomalous
	X-ray pulsars: magnetar candidates. in {\it{Compact Stellar X-ray
 	Sources}} (eds. Lewin, W.~H.~G. \& van der Klis, M.), in press
 	(astro-ph/0406133) (2005).
\bibitem{zhaoetal92} Zhao, J-H. {\it{et al.}} A transient radio source near the	
	center of the milky way galaxy. {\it{Science}} {\bf{255}}, 1538-1543 (1992).


\end{thebibliography}
\end{document}